\documentstyle[11pt,aaspp4]{article}

\tightenlines

\lefthead{Risaliti, Maiolino \& Salvati}
\righthead{Distribution of N$_H$ among Seyfert 2s}

\begin{document}

\title{The distribution of absorbing column densities \\
    among Seyfert 2 galaxies}

\author{G. Risaliti}
\affil{Dipartimento di Astronomia e Scienza dello Spazio, Univerit\`a di
Firenze, L. E. Fermi 5, I-50125, Firenze, Italy}

\author{R. Maiolino and M. Salvati}
\affil{Osservatorio Astrofisico di Arcetri, L. E. Fermi 5, 
I-50125 Firenze, Italy}


\begin{abstract}

We use hard X-ray data for an ``optimal'' sample of Seyfert 2 galaxies
to derive the distribution of the gaseous absorbing column
densities among obscured active nuclei in the local Universe.
Of all Seyfert~2 galaxies in the sample, 75\%  are heavily obscured  (N$_H > 10^{23} \rm
cm^{-2}$) and about half are Compton thick (N$_H > 10^{24} \rm cm^{-2}$).
Intermediate type 1.8--1.9 Seyferts are characterized by an average
N$_H$ much lower than ``strict'' Seyfert 2s. No correlation is 
found between N$_H$ and the intrinsic luminosity of the nuclear source.
This N$_H$ distribution has important consequences for the synthesis of the
cosmic X-ray background. Also, the large fraction of Compton thick objects 
implies that most of the obscuring gas is located within a radius of a 
few 10 pc from the nucleus.

\end{abstract}


\section{Introduction}

According to the so-called unified model (Anto\-nuc\-ci 1993) the same
engine
is at work in all Active Galactic Nuclei (AGNs). The
differences
between type 1 and type 2
AGNs are ascribed solely to orientation effects: our line of sight to
the
nucleus may (type 2) or may not (type 1) be obstructed by optically
thick
material, perhaps distributed in a toroidal geometry.
The knowledge of the amount of obscuring gas in Seyfert galaxies is
important
to understand the physical properties and the nature of the putative
torus, but is also relevant to other AGN--\-related issues, such as the
cosmic X-ray background. Indeed, the distribution of the absorbing N$_H$
is a key ingredient in background synthesis models
(Comastri et al. 1995, Gilli et al. 1999a).

When the photoelectric absorption cutoff is observed in the hard X-ray
spectrum of a given source, the column density of the obscuring gas is
easily measured, at least in principle. In the spectral analysis of these
sources an extra emission at low energies is often present, the so-called
``soft excess'' (ascribed either to scattered radiation or to diffuse thermal
emission), which is fitted with a separate model component. The N$_H$ which
is derived from the fit is the one that obscures the central emission of
the AGN, while other opacities affecting the reflected or diffuse radiation 
are not relevant in the present context.
When instead only a reflected component is visible, and the direct 
component is completely suppressed in the observed spectral range,
one can deduce only a lower limit for N$_H$, which
depends on the maximum energy for which measurements are available (for
a detailed discussion see Maiolino et al. 1998a, hereafter M98). 

Many Seyfert 2s (hereafter Sy2s)
were observed in the past by various X-ray satellites such
as Einstein, Ginga and, more recently, ASCA, ROSAT, and BeppoSAX.
The main limitation of previous studies on the N$_H$ distribution of
Sy2s was the strong bias in favor of X-ray
bright sources, which tend to be the least absorbed ones.
However, recent observations have probed X-ray weaker Sy2s,
partly removing the selection against heavily obscured objects (Salvati et al.
1997, Awaki et al. 1997, M98, Risaliti et al. 1999).
Several works have tackled the issue of the N$_H$ distribution by taking
advantage of these new data.
M98 compare the N$_H$ distribution of past surveys with that obtained from
the BeppoSAX survey of an [OIII]--selected Sy2 sample, and emphasize the bias
against heavily absorbed objects in the former. Bassani et al.
(1999, hereafter B99) present the N$_H$ distribution of the
whole sample of Sy2s for which hard X-ray data are available, and of an
optically selected subsample. All of these
works are still affected by selection effects or by incompleteness problems;
they are nonetheless useful to monitor how the selection criteria can affect 
the shape of the N$_H$ distribution. 

Here we start from an optical sample of Seyfert 2 galaxies which is the best
available one in terms of biases related to absorption. From it we
extract a nuclear--flux limited subsample for which almost complete hard
X-ray information exists, either published, or produced by us from
archival data. The absorbing column density distribution is derived
from this ``optimal'' subsample.
In Section 2 we describe the selection criteria of the subsample, and
discuss the correction of residual biases and incompleteness effects.
In Section 3 we present the N$_H$ distribution which is obtained, and
briefly analyze the physical consequences of our results.

\section{Sample selection.}

In this paper with the name ``Seyfert 2'' we refer to sources classified
as type 1.8, 1.9, and 2 in optical catalogs. Our ``parent''
sample of Seyfert 2 galaxies is that of Maiolino \& Rieke
(1995, hereafter MR), completed with NGC 1808\footnote{This source fits the
selection criteria of Maiolino \& Rieke, but was missed by the original version
of the sample.}. We also consider 18 new Seyferts found by
Ho et al. (1997), which would have been included in MR if they had been
discovered earlier. These 18 additional sources are useful in order to test MR
in terms of completeness and residual biases, as will become clear in the
following; however, none of them will be included in the final subsample
(because of the selection criteria of the latter), and we do not have
to tackle problems related to possible dishomogeneities between MR and 
Ho et al. The MR Seyferts are identified
spectroscopically within the Revised Shapley Ames catalog of galaxies
(RSA, Sandage \& Tammann 1987), which is limited in the B magnitude
of the host galaxy (B$_T < 13.4$ mag). Ho et al. (1997) select their Seyfert
galaxies with the same criterion, but they cut the RSA to a lower limiting
magnitude (B$_T < 12.5$ mag) and require that $\delta > 0 ^{\circ}$;
on the other hand, they generally have spectra with higher signal-to-noise.
Hard X-ray spectra are available for 43 of the MR Sy2s and for 1 out of the
18 additional sources of Ho et al.
The naive approach of taking the N$_H$ distribution of these 44 objects
cannot be adopted, however, because of possible
residual biases and incompleteness problems.
We discuss these issues in the following.

As mentioned in the Introduction, a large fraction of the Sy2s with hard
X-ray spectra were observed because they were known to be bright according
to all--sky X-ray surveys, and
very likely this selection criterion introduced a bias in
favor of little absorbed objects (see M98 for a detailed
discussion). A Sy2s (sub)sample suitable for determining a more reliable
distribution of N$_H$ should be selected according to criteria
independent of absorption effects. The optimal sample should be limited in
the {\it intrinsic flux} (i.e. before absorption) of the active nucleus.

A good indicator of the intrinsic
AGN flux is the intensity of the [OIII] $\lambda 5007$~\AA\ emission
line that is produced in the Narrow Line Region (NLR) on
the 100--pc scale and, therefore, is little
affected by the nuclear obscuration on the pc scale.
However, MR
showed that the host galaxy gaseous disk might obscure part of the
NLR (see also di Serego Aligheri et al. 1997 and Hes et al. 1993), and it
is important to correct the [OIII] flux for large scale absorption.
The magnitude of
the correction can be derived from the (narrow line) Balmer decrement.
According to the unified model, the ratio between the
{\it observed} and the {\it intrinsic} X-ray luminosity of a Seyfert
nucleus
is a measure of the absorbing N$_H$ along the line of sight.
Therefore, if the (extinction corrected) [OIII] luminosity is proportional
to the intrinsic luminosity, then an estimate of
N$_H$ is provided by the ratio between the observed
X-ray flux and the corrected [OIII] flux.
The expected relation was confirmed and calibrated by a recent
statistical study on a large sample of Seyfert 2 galaxies (B99).

\placetable{t1}

\placetable{t2}

In Table 1 we report the MR sample in order of decreasing
absorption--corrected [OIII] flux, while in Table 2 we report the 18 additional
sources from the Ho et al. sample. All data are taken from the
literature, with the exception of
5 objects (NGC 4565, NGC 5005, IC 2560, NGC 5347,
and IC 5135), which have been observed but whose hard X-ray data have
not been published yet. For these sources, we have retrieved and
analyzed the data from the ASCA public archive. The results of our
analysis are briefly summarized in the Appendix. NGC 4945 and IC 2560 do not
have [OIII] data, but their [OIII] flux is inferred to be quite high
based on their X-ray flux (see the Appendix for more details).
As discussed above, most of the sources in Table 1
were selected for hard X-ray observation because
they were found bright in previous all--sky X-ray surveys.
Only 13 of them were selected because of their strong [OIII] flux
($> 40\times 10^{-14}$ erg cm$^{-2}$ s$^{-1}$) without
a previous X-ray survey detection
(Salvati et al. 1997, M98, Risaliti et al. 1999, Awaki et al. 1997).
The N$_H$ distribution of the latter set is strongly biased
towards high values of N$_H$ (M98), as expected, since their
non--detection in X-ray surveys, and their high intrinsic flux implied by 
the [OIII] can be reconciled only by a strong pc--scale absorption.

We now note that Table 1 has 9 sources with ``high'' [OIII] flux
($ > 40\times 10^{-14}$ erg s$^{-1}$ cm$^{-2}$) and with no hard X-ray
data, or with a signal-to-noise not good enough to provide spectroscopic
information\footnote{This is the case of NGC 1667,
which shows a decreasing flux in a timescale of years, so that
the N$_H$ is unknown (see B99 for more details).}.
To all these 9 sources one can apply the same line of reasoning of the 13
[OIII]--selected sources mentioned above, and one can plausibly assume
that both sets have the same distribution of column densities.

The 18 additional sources discovered in the RSA by Ho et al. (1997) have
systematically
low [OIII] fluxes ( F$_{[OIII]} = 33\times 10^{-14}$ erg s$^{-1}$
cm$^{-2}$ for the brightest one, the others have F$_{[OIII]}< 20\times
10^{-14}$ erg cm$^{-2}$s$^{-1}$):
they are intrinsically weak, in accordance with
their being unnoticed until recently in the optical. These new additions
to the original MR sample are evidence that the latter is incomplete at
low [OIII] fluxes, and the incompleteness could entail a bias in favor
of low absorption in low luminosity objects. This would be analogous to
the bias that affected older optically selected Seyfert samples which
indeed, as shown by MR, avoided low luminosity, highly obscured AGNs.
The same bias might show up here, since these faint nuclei are in any
case difficult to detect and a substantial absorption could take them below
the detection limit.

\placefigure{f1}

Hints of this effect at low fluxes come from the comparison of the
N$_H$ values of the F$_{[OIII]} > 40 \times 10^{-14}$ erg cm$^{-2}$
s$^{-1}$ subsample with the one having $3 \times 10^{-14}$ erg cm$^{-2}$ 
s$^{-1} <$ F$_{[OIII]}  < 40 \times 10^{-14}$ erg cm$^{-2}$s$^{-1}$ (there 
are no X-ray data for sources with F$_{[OIII]} < 3 \times 10^{-14}$ erg 
cm$^{-2}$s$^{-1}$; here, only in order to compare high and low flux 
sources, we merge together Tabs. 1 and 2). Out of the 45 brightest Sy2s
only 2 objects (i.e. 5\%) have N$_H<10^{22}$ cm$^{-2}$, while among 
the next fainter 27 at least 6 (i.e. 22\%) have such a low N$_H$.
Assuming that the true distribution
is given by the first set, there is already a 2 $\sigma$ excess of low absorption objects
in the second set, notwithstanding the large incompleteness of its hard
X-ray data.

\placefigure{f2}

Alternately, the differences between the two N$_H$ distributions could
arise because of a real physical dependence of the column
density on the intrinsic luminosity of the source.
In order to explore this hypothesis, we looked for a correlation between
N$_H$ and [OIII] luminosity among the 45 strongest [OIII]
sources, and found none (upper panel of Fig. 1).
The same result (i.e. no correlation) is
found by comparing the N$_H$ distribution directly with the absorption
corrected X-ray (2--10 keV)
luminosity\footnote{This can be done only for objects
whose X-ray spectrum shows the direct transmitted component, not for
those which are completely Compton thick.}, as shown 
in the lower panel of Fig. 1.

Further evidence for a bias affecting the [OIII]--faint nuclei
comes from the axial ratios (a/b) of the host galaxies (Fig. 2):
the faintest [OIII] sources are seen preferentially face--on, while
the brightest ones have an isotropic distribution
for a/b $>$ 0.2 (axial ratios lower than 0.2 are difficult to find because 
of the galactic disk thickness). The reason is probably that
faint AGNs are hardly detected if the host galaxy
is edge--on, since then the NLR emission is partially absorbed.
This result further supports the idea that the subsample of faint objects
is biased in terms of obscuration.
To prevent our results from being affected by the same bias, we excluded
from our study all the objects with an
[OIII] flux less than 40 $\times 10^{-14}$ erg cm$^{-2}$s$^{-1}$.
This selection criterion automatically excludes all the additional 
sources from the Ho et al. sample; this is illustrated in Fig. 3 where
the [OIII] flux distribution is shown.

\placefigure{f3}

With the sample selection discussed above,
we do not use all the available data (8 of the
ignored objects have X-ray spectra: 18\% of the total), but we can
assume that the remaining 45 objects constitute the least biased
subsample now available.

A final concern regards the optical classification of some of the objects in
Tab. 1, that might be different depending on the reference in the literature,
and might pose into question whether these objects should be included or not.
Some of the discrepant classifications are simply a matter of
sensitivity of the optical spectra, and affect mostly faint objects, i.e. 
those with low [OIII] fluxes, that we have excluded anyway.
In some intermediate type Seyferts (1.8--1.9) the slit
width plays a role: narrower slits tend to miss a fraction of the 
NLR thus turning some of the nearby Sy1.8s into Sy1.5s; conversely, 
missing the nucleus with the slit could cause the opposite error.
To minimize this problem we adopted the classification obtained with the
widest slit, as long as the signal-to-noise of the spectra were comparable.
Again, most of these intermediate Seyferts with discrepant classification
are in the low--[OIII] part of Tab. 1 that we have excluded. Only two of them
are in the high--[OIII] subsample, namely NGC 1275 and NGC 3031 (classified as
Sy1.5s by Ho et al.): these two objects have low N$_H$, therefore their
exclusion would further shift the N$_H$ distribution towards high values. 
Variability can also play a role: in objects known to be variable (e.g.
NGC 2992) we adopted the classification and the N$_H$ measurement in the high
state. Finally, a few objects have line ratios at the borderline with
LINERs, whose relationship with Seyfert nuclei is still matter of debate.
Only one object of this category is present in the high--[OIII] subsample,
namely NGC 5005. Summarizing, uncertainties in the optical classification
do not affect significantly our results thanks to our [OIII]--based
selection criterion: at most 3 out of the 45 objects in the final subsample might
be misclassified.

\placefigure{f4}

\section{The N$_H$ distribution and its implications.}

Our final subsample is composed of 45 sources: we have direct N$_H$
measurements for 36 of them, while for the remaining 9 we can make
reasonable assumptions based on their ``detection history'' in optical
and X~rays (see previous Section).
The results are plotted in Fig. 4, and tabulated in Table~3;
the number in brackets corresponds to objects for which only a
lower limit on N$_H$ is available. Quoted errors are 1~$\sigma$, and they are
estimated by means of a Monte Carlo simulation in two steps: first, 9 N$_H$ 
values are extracted at random according to the distribution of
the 13 [OIII]--selected sources mentioned in Sect.~2; these are added to the 
36 known values in order to obtain a 45--entry distribution. Then, 45 N$_H$ 
values are extracted at random according to the latter distribution. The 
procedure is repeated many times and the 1--$\sigma$ interval for each bin is 
measured.

In the hypothesis that the [OIII] flux is a good indicator of the
AGN intrinsic luminosity, and that N$_H$ is independent of luminosity,
every [OIII] selected subsample should provide the same
N$_H$ distribution. We can therefore check
our assumptions about the 9 objects not observed in the X rays
by repeating our analysis on a smaller set, composed only by the
first 30 objects in Tab. 1 with available hard X-ray data (3 sources
without such data are interspersed among them, and are ignored).
The new N$_H$ distribution is in
agreement, within the errors, with the one plotted in Fig. 4. Obviously
the relative errors are higher, the statistics being lower, but the check
confirms that our treatment of the 9 non-observed sources
does not introduce new biases.

\placetable{t3}

The shape of the distribution is quite different from what was assumed
up to now: we obtain that $\sim \frac{3}{4}$ of all
Sy2s are heavily obscured (N$_H > 10^{23}$ cm$^{-2}$) and about half
are Compton thick (N$_H > 10^{24}$ cm$^{-2}$).
The main uncertainty still remaining is about the fraction of completely
opaque sources (N$_H > 10^{25}$ cm$^{-2}$) with respect to
``translucent''
objects with $10^{24}$ cm$^{-2} < N_H < 10^{25}$ cm$^{-2}$.
In both cases the spectra are reflection dominated in the 2--10
keV band, and cannot be distinguished by ASCA or
other medium X-ray satellites. The only way of identifying column
densities
in the interval $10^{24}$ cm$^{-2} < N_H < 10^{25}$ cm$^{-2}$ is by
means of
solid scintillating detectors sensitive around 30~keV, like the PDS
instrument
on board BeppoSAX.

The shaded bar in Fig. 4 indicates sources with reflection dominated
spectra in the 2--10 keV spectral range, but not observed at higher energies,
i.e. objects for
which only a lower limit of 10$^{24}$ cm$^{-2}$ can be set to the absorbing
column density.
Placing them in the $10^{24}$ cm$^{-2} < N_H < 10^{25}$ cm$^{-2}$ bin
is a conservative assumption, which minimizes the amount of absorption
required; also, a substantial population in this bin is needed for the
synthesis of the X-ray background (Gilli et al. 1999b). However,
if we consider the objects with sensitive observations in the 10--200 keV
range (mostly by BeppoSAX), an unexpected low number of
sources with $10^{24}$ cm$^{-2} < N_H < 10^{25}$ cm$^{-2}$ comes out. Up
to now, 10 Seyfert 2s with N$_H > 10^{24}$ cm$^{-2}$ were observed:
only 2 of them have also N$_H < 10^{25}$ cm$^{-2}$, 7 have N$_H > 10^{25}$
cm$^{-2}$, while for the remaining one the estimate is controversial.
Future observations, to be performed by means of BeppoSAX and the
next Japanese X-ray satellite Astro--E, will make clear whether the
present lack is due to statistical fluctuations or not.

As outlined previously, our sample of 45 sources is composed by type
1.8, 1.9 and 2 Seyfert galaxies, i.e. sources that generally show
indications of cold absorption in excess of the Galactic value.
Nevertheless, it can be interesting to consider the N$_H$ distribution of 
type 2 and type 1.8--1.9 objects separately, as done in Fig. 5.
The average column density of the intermediate Seyferts is clearly
much lower than that of the ``strict'' Sy2s, in
agreement with the expectations of the model proposed
by Maiolino \& Rieke (1995).

\placefigure{f5}

The results reported here have several astrophysical implications. Present
models of the cosmic X-ray background (Comastri et al. 1995, Gilli et
al. 1999a) are able to reproduce the observations at energies between 2 keV
and 100 keV in terms of the contribution of the AGNs (both type 1 and type 2).
Among the many parameters of such models (AGN luminosity function and
evolution, type~1 to type~2 ratio, components of the mean
spectrum)
only the ones concerning the N$_H$ distribution were not constrained by
any
direct measurement until now.
The inclusion of our results on the N$_H$ distribution
in the synthesis codes, together with
new determinations of the X-ray luminosity function and evolution of type~1
AGNs (Miyaji et al. 1999), leads to very important conclusions on the
properties of the contributors to the hard X-ray background and, specifically,
on the existence of high luminosity type~2 AGNs (QSO2s),
the evolution with redshift of
the type~1 to type 2~ratio, and the consistency with the hard X-ray
counts
which are now becoming available (Giommi et al. 1998, Cagnoni et al.
1998).
All these issues will be tackled in a forthcoming paper (Gilli et al. 1999b).

Our results on the distribution of the absorbing N$_H$ has important implications
also on the spatial extension of the obscuring medium. Indeed, it is 
not clear yet whether the obscuring gas is confined in a relatively
dense circumnuclear torus (with typical radius of a few pc), or is
distributed in a larger volume on the 100-pc--1-kpc scale.
Obscuring material distributed on the 100--pc scales has been observed
directly by means of HST images of nearby AGNs (e.g. Malkan et al. 1998);
it has been inferred also from the shape of the infrared spectra
(Granato et al. 1997) and from the shortage of edge--on systems in several
Seyfert galaxy samples (Keel 1980,
Maiolino \& Rieke 1995, Simcoe et al. 1997). If we assume that the
obscuring gas is distributed within a radius $R$ from the nucleus and
with a covering factor $f$, then the gas mass enclosed within the region
is
\begin{equation}
\left( {M_{gas}\over M_{\odot}} \right) \simeq 3.5\times 10^8
 \cdot f \cdot \left( {N_H \over 10^{24}\rm cm^{-2}} \right)
 \left( {R\over \rm 100~pc}\right) ^2
\end{equation}
where N$_H$ is the absorbing column density along the lines of sight that
intersect the obscuring medium\footnote{Eq. 1 is roughly independent of the
assumed geometry, unless most of the absorbing gas is distributed in a thin
shell whose thickness is much smaller than its radius. As we shall see, the
latter distribution is very unlikely, since the covering factor 
is larger than 50\%, and a thin shell so much spread out would be problematic 
on grounds of dynamical stability and physical origin.}. If we assume
that the covering factor $f$ is responsible for the observed
Sy2/Sy1 ratio ($\simeq 4$)\footnote{Both MR and Ho et al. (1997) give
Sy2/Sy1 $\approx 4$ if Sy1.8s and Sy1.9s are grouped together in the type 2
class.} and for the
opening angle of the observed light cones, then $f\sim 0.8$. In the case
of Compton thick Sy2s a radius of the obscuring region of the order of
100 pc would imply a gas mass of the order of or
larger than $10^9~ \rm M_{\odot}$, that in many objects exceeds by itself
the dynamical mass in the same region. One way to have
the Compton thick obscuring gas extended over the 100--pc scale would
be to assume that its covering factor $f_{CT}$ is very small ($\ll 0.1$). 
Until a few years ago, when the only
known Compton thick source was NGC 1068, this possibility could have been
plausible. However, our new results imply $f_{CT}\sim 0.4$. As a consequence 
in most AGNs the Compton thick gas must be distributed on scales significantly
smaller than 100 pc. We can investigate this issue in detail in two well 
studied sources, Circinus and NGC 1068. Circinus has an N$_H = 4.3\times 
10^{24} \rm cm^{-2}$ (Matt et al. 1999),
and Maiolino et al. (1998b) have mapped its dynamical mass by means of
integral field spectroscopy on scales from 100 pc down to the central 10 pc.
They concluded that the Compton thick gas must be contained within $R <
20$  pc not to exceed the dynamical mass. As for NGC 1068, if the Compton 
thick gas (N$_H >10^{25} \rm cm^{-2}$)
were distributed over the central 200 pc then the implied
gas mass ($> 1.6\times 10^9~ \rm M_{\odot}$)
would exceed the dynamical mass by a factor larger
than 3 (Thatte et al. 1997), and would exceed
the gas mass estimated by means of CO observations
by a factor larger than 100 (Helfer 1997). On the other hand, gas 
with lower column densities ($\sim 10^{23}\rm cm^{-2}$ or lower) could extend
over the 100--pc scale.

\section{Summary}

We combine a large number of published and archived hard X-ray data
of Seyfert 2 galaxies in the Maiolino \& Rieke (1995) sample,
that provide information on the absorbing column density
N$_H$. From this database we select
a subsample of objects that is limited in the {\it intrinsic} flux of
the active nucleus, as deduced from the reddening corrected [OIII] line.
After correcting for a residual incompleteness, we derive
the distribution of the absorbing N$_H$ of this subsample. This can be 
regarded as the best approximation (to date) to the real distribution of 
N$_H$ for the local population of (moderately luminous) Seyfert 2 nuclei.
Statistical errors of the distribution are also derived.

The N$_H$ distribution turns out to be quite different from
previous estimates. In particular, 75\% of all Sy2s in our final sample
are heavily obscured
(N$_H > 10^{23}~ \rm cm^{-2}$) and about half are Compton thick
(N$_H > 10^{24}~ \rm cm^{-2}$).

If treated separately, intermediate type 1.8--1.9 Seyfert galaxies are
characterized by
an average N$_H$ much lower than ``strict'' Seyfert 2s. 

In previous models of the cosmic
X-ray background the largest set of free parameters was
contained in the N$_H$ distribution. Our empirical distribution 
can now be used instead, and
it is expected to pose tight constraints to these models.

The N$_H$ distribution has also important implications on the spatial
extension of the absorbing medium (the putative torus). The large fraction of
Compton thick Seyfert 2s implies that most of the Compton thick gas must
be located within a few 10 pc from the nucleus, in order not to exceed the
dynamical mass in the central region. However, gas with low N$_H$ could
extend to larger radii (100 pc or more).

Finally,
we do not find evidence for any correlation between the absorbing N$_H$
and the intrinsic luminosity of active nuclei, over a range of luminosities
spanning 3 orders of magnitude.

\acknowledgments

We are grateful to S. Ueno for providing us with information on his
data in advance of publication. We thank the anonymous referee for
helpful comments.
This work was partially supported by the Italian Space Agency (ASI)
through the grant ARS--98--116/22.

\appendix
\section{Notes on individual objects}

NGC 4945 and IC 2560 do not have [OIII] data, but their X-ray properties
imply that the intrinsic [OIII] flux must be high based on the following
considerations. NGC 4945 is one of the nearest active galaxies and
is among the brightest AGNs at energies around 100 keV.
The absence of [OIII] data cannot be ascribed to an intrinsic weakness
of the source, but only to the very high extinction in
the edge--on galactic disk, and F$_{[OIII]} > 40 \times 10^{-14}$ erg
s$^{-1}$ cm$^{-2}$ is a conservative assumption. IC 2560 has an X-ray flux of
4.8$\times 10^{-13}$ erg s$^{-1}$ cm$^{-2}$, and is certainly Compton thick
(see below). If F$_{[OIII]} \le 40 \times 10^{-14}$ erg cm$^{-2}$s$^{-1}$
cm$^{-2}$, the F$_{2-10 keV}$/F$_{[OIII]}$ ratio would be higher than 1. On the
contrary, the diagnostic diagram for Seyfert 2s in B99 strongly suggests that
this ratio is significantly lower than 1 in Compton thick sources, so very likely
F$_{[OIII]} > 40 \times 10^{-14}$ erg s$^{-1}$ cm$^{-2}$.

Finally we report the results of the data analysis of 5 Sey\-fert~2s taken from
the ASCA public archive. All errors are at the 90\% confidence level
for one interesting parameter.

\begin{enumerate}
\item{} NGC 4565 has a powerlaw spectrum with spectral index
1.9$^{+0.45}_{-0.30}$ and no evidence of absorption.
The 2--10 keV flux is 1.8$\times 10^{-12}$ erg cm$^{-2}$s$^{-1}$.
Both the spectral shape and the high X/[OIII] flux ratio
(F$_{2-10 keV}$/F$_{[OIII]}$ = 30)
strongly suggest that this is a Compton thin source.
The spectral fit constrains the N$_H$ to be lower than $6\times10^{21}~
\rm cm^{-2}$ (at a confidence level of 90\%).

\item{} NGC 5005 shows a steep powerlaw spectrum with spectral index
2.15$^{+0.35}_{-0.30}$ and no absorption. The 2--10 keV flux is 3$\times
10^{-13}$ cm$^{-2}$erg s$^{-1}$ and the 6.4 keV iron K$_\alpha$ line is not
detected, but with a high upper limit on its equivalent width (900 eV).
The X/[OIII] flux ratio is low
(F$_{2-10 keV}$/F$_{[OIII]}$ = 0.15), locating this source among the Compton
thick sources in the diagram of B99.
 There is no evidence of a cold reflected component at high
energies (E$>$5 keV). Our interpretation is that we see only
the extended starburst component of the host galaxy, while the AGN
X-ray emission is fully absorbed. The absence of the reflected
component may be due to orientation effects (the reflecting torus
may be almost perfectly edge--on).
For this reason we estimate a lower limit
for the nuclear N$_H$ of 10$^{24}$ cm$^{-2}$, which is the minimum
value required to absorb all the direct emission in the ASCA
spectral range.

\item{} IC 2560 has a prominent iron
line with EW(Fe)=6.3$_{-3.0}^{+2.6}$ keV (E=6.56$^{+0.25}_{-0.15}$ keV)
typical of a Compton thick Sy2. Since it has been observed only in the 2--10
keV band,
we can only set a lower limit of 10$^{24}$ cm$^{-2}$ to N$_H$.
The 2--10 keV observed flux is 4.8$\times 10^{-13}$ erg s$^{-1}$ cm$^{-2}$.

\item{} NGC 5347 is another Compton thick source, with EW(Fe)$>$1.9 keV and
F(2--10keV)=3.8$\times 10^{-13}$ erg cm$^{-2}$ s$^{-1}$. Therefore,
we can put a lower limit of 10$^{24}$ cm$^{-2}$ to N$_H$.

\item{} IC 5135 has a low signal to noise spectrum (the detection is at
5.5$\sigma$), from which we can only estimate the
flux F(2--10keV)=5.1$\times 10^{-13}$ erg cm$^{-2}$ s$^{-1}$ and a flat
spectral index ($\Gamma < 1.40$). The Fe K$_{\alpha}$ line is detected
with an equivalent width of 1.37$^{+1.2}_{-1.2}$ keV. The low value of the
photon index, the large value of the EW(Fe) and
the low X/[OIII] ratio (F$_{2-10 keV}$/F$_{[OIII]}$=0.5) strongly
support the idea that this is a Compton reflection dominated spectrum,
so N$_H > 10^{24}$ cm$^{-2}$ for this source as well.
\end{enumerate}

\clearpage
 
 \begin{deluxetable}{cccc|cccc}
 \footnotesize
 \tablecaption{The MR sample of Seyfert 2 galaxies\label{t1}}
 \tablewidth{0pt}
 \tablehead{
 \colhead{} & \colhead{Name}   &
 \colhead{F$_{[OIII]}$\tablenotemark{\dagger}}
 & \colhead{N$_H$ \tablenotemark{\ddagger}} & \colhead{} &
 \colhead{Name} &
 \colhead{F$_{[OIII]}$\tablenotemark{\dagger}} &
 \colhead{N$_H$ \tablenotemark{\ddagger}} \nl
 }
 \startdata
 1&NGC1068&15800&$>10^5$(a) &
 38&NGC 3185&74&\nl
 2&CIRCINUS&6970&43000$^{+19000}_{-11000}$(b)&
 39&NGC 6221&72& \nl
 3&IRS 07145&2010&$>10^5$(a) &
 40&NGC 5674&59&700$^{+280}_{-260}$(a)\nl
 4&NGC 7314&1770&116$^{+4}_{-13}$(a)&
 41&NGC 1320&57& \nl
 5&IRS 18325&752&132$^{+10}_{-10}$(a)&
 42&NGC 3281&45&7980$^{+1900}_{-1500}$(a)\nl
 6&NGC 5643&694&$>10^5$(a) &
 43&NGC 3031&43&9.4$^{+0.7}_{-0.6}$(a)\nl
 7&NGC 2992&680&69$^{+33}_{-19}$(a)&
 44&NGC 4945& $>40^{**}$ &40000$^{+2000}_{-1200}$(a)\nl
 8&NGC 1386&655&$>10^4$(a)&
 45&IC 2560&$>40^{**}$ &$>10^4$ (e) \nl
 \cline{5-8}
 9&IC 3639&620&$>10^5$(c) &
 46&NGC 7743&40& $\uparrow$ sample limit \nl
 10&NGC 5135&614&$>10^4$(a) &
 47&NGC 4501&36& \nl
 11&NGC 5506&600&340$^{+26}_{-12}$(a)&
 48&NGC 7496&29& \nl
 12&MKN 1066&513&$>10^4(d)$ &
 49&NGC 788& 27& \nl
 13&NGC 7582&445&1240$^{+60}_{-80}$(a)&
 50&NGC 6890&25&\nl
 14&NGC 4388&374&4200$^{+600}_{-1000}$(a)&
 51&NGC 6300&20&\nl
 15&NGC 4941&355&4500$^{+2500}_{-1400}$(a)&
 52&NGC 4395&19&\nl
 16&IC 5063&353&2400$^{+200}_{-200}$(a)&
 53&NGC 1358&19&\nl
 17&NGC 2110&321&289$^{+21}_{-29}$(a)&
 54&NGC 7590&17&$<9.2$(a)\nl
 18&NGC 3393&316&$>10^5$ (a)&
 55&NGC 5033&17&8.7$^{+1.7}_{-1.7}$(a)\nl
 19&NGC 1275&311&149$^{+69}_{-69}$(a)&
 56&NGC 7410&16&\nl
 20&NGC 2273&277&$>10^5$ (a)&
 57&NGC 7479&16&\nl
 21&NGC 4258&262&1500$^{+200}_{-200}$(a)&
 58&IRS 11215&11&\nl
 22&NGC 5194&228&7500$^{+2500}_{-2500}$(a)&
 59&NGC 4579&9&4.1$^{+2.7}_{-2.7}$(a)\nl
 23&NGC 3081&215&6600$^{+1800}_{-1600}$(a)&
 60&NGC 3147&9&4.3$^{+3.2}_{-2.7}$(a)\nl
 24&NGC 5005&202&$>10^4$(e)&
 61&NGC 3786&8 &\nl
 25&NGC 1667&197&&
 62&NGC 4594&7&55$^{+41}_{-40}$(a)\nl
 26&MKN 1073&195&&
 63&NGC 5128&6&1000-3500(a)\nl
 27&NGC 5728&180&&
 64&NGC 5427&$>6^{\ast}$&\nl
 28&NGC 4507&158&2920$^{+230}_{-230}$(a)&
 65&NGC 1241&$>4^{\ast}$&\nl
 29&NGC 1365&141&2000$^{+400}_{-400}$(a)&
 66&NGC 7172&4&861$^{+79}_{-33}$(a)\nl
 30&NGC 1808&131&320$^{+588}_{-318}$(a)&
 67&NGC 2639&4 &\nl
 31&NGC 4939&112&$>10^5$ (a)&
 68&NGC 5273&3&\nl
 32&NGC 5347&100&$>10^4$ (e)&
 69&NGC 1433&2&\nl
 33&IC 5135&97&$>10^4$ (e)&
 70&NGC 4639&0.9 &\nl
 34&NGC 3982&90& &
 71&NGC 513& - &\nl
 35&NGC 3079&90&160$^{+270}_{-130}$(a)&
 72&NGC 4785&- & \nl
 36&NGC 5953&86& &
 73&NGC 3362& - &\nl
 37&NGC 5899&74& &
 74&NGC 7465& - &\nl
  
\enddata
   
\tablenotetext{\dagger}{[OIII] flux in units of 10$^{-14}$ erg
s$^{-1}$
cm$^{-2}$, corrected for the
extinction as deduced from the (narrow line)
Balmer decrement. No data are available for the last four
sources.}
\tablenotetext{\ddagger}{N$_H$ in units of 10$^{20}$ cm$^{-2}$.}
\tablenotetext{\ast}{Not corrected for the Balmer decrement.}
\tablenotetext{**}{Estimated (see text).}
\tablenotetext{}{References: (a) B99, (b) Matt et al. (1999),
(c) Risaliti et al. 1999,
 (d) Ueno private communication, (e) this work.}
  
\end{deluxetable}
   
\clearpage
    
\begin{deluxetable}{cccc|cccc}
\footnotesize
\tablecaption{Additional Sy2s from the Ho et al. sample.
\label{t2}}
\tablewidth{0pt}
\tablehead{
\colhead{} & \colhead{Name}   &
\colhead{F$_{[OIII]}$\tablenotemark{\dagger}}
& \colhead{N$_H$ \tablenotemark{\ddagger}} & \colhead{} &
\colhead{Name} &
\colhead{F$_{[OIII]}$\tablenotemark{\dagger}} &
\colhead{N$_H$ \tablenotemark{\ddagger}} \nl
}
\startdata
 
 1&NGC 3735&33&&         10&NGC 3976&2.4&\nl
 2&NGC 1167&17&&         11&NGC 4698&2.4&\nl
 3&NGC 2655&16&&         12&NGC 4725&2.1&\nl
 4&NGC 4565&6&$<60$ (a)& 13&NGC 3486&1.7&\nl
 5&NGC 3254&5&&          14&NGC 1058&1.3&\nl
 6&NGC 4477&4&&          15&NGC 4378&0.8&\nl
 7&NGC 3941&3&&          16&NGC 185&0.2&\nl
 8&NGC 4138&2.9&&        17&NGC4472&0.1&\nl
 9&NGC 676&2.9&&         18&NGC 4168&0.1&\nl
      
\enddata
       
\tablenotetext{\dagger}{[OIII] flux in units of 10$^{-14}$ erg
s$^{-1}$
cm$^{-2}$, corrected for the
extinction as deduced from the (narrow line)
Balmer decrement.}
\tablenotetext{\ddagger}{N$_H$ in units of 10$^{20}$ cm$^{-2}$.}
\tablenotetext{}{References: (a) this work}

\end{deluxetable}

\clearpage

\begin{table}
\begin{center}
\begin{tabular}{c|c}
log(N$_H$)(cm$^{-2}$)& \% Sy2s\\
\hline
$<$ 22 & 4.44$\pm2.87$ \\
22-23 & 17.78$\pm5.64$ \\
23-24 & 29.91$\pm7.72$ \\
24-25 & 21.53$\pm7.03$ (17.09) \\
$>$25   & 26.33$\pm7.13$ \\
\end{tabular}
\caption{The numerical values corresponding to Figure~4.}
\label{t3}
\end{center}
\end{table}


\clearpage

{\bf Figure captions:}\\

\figcaption[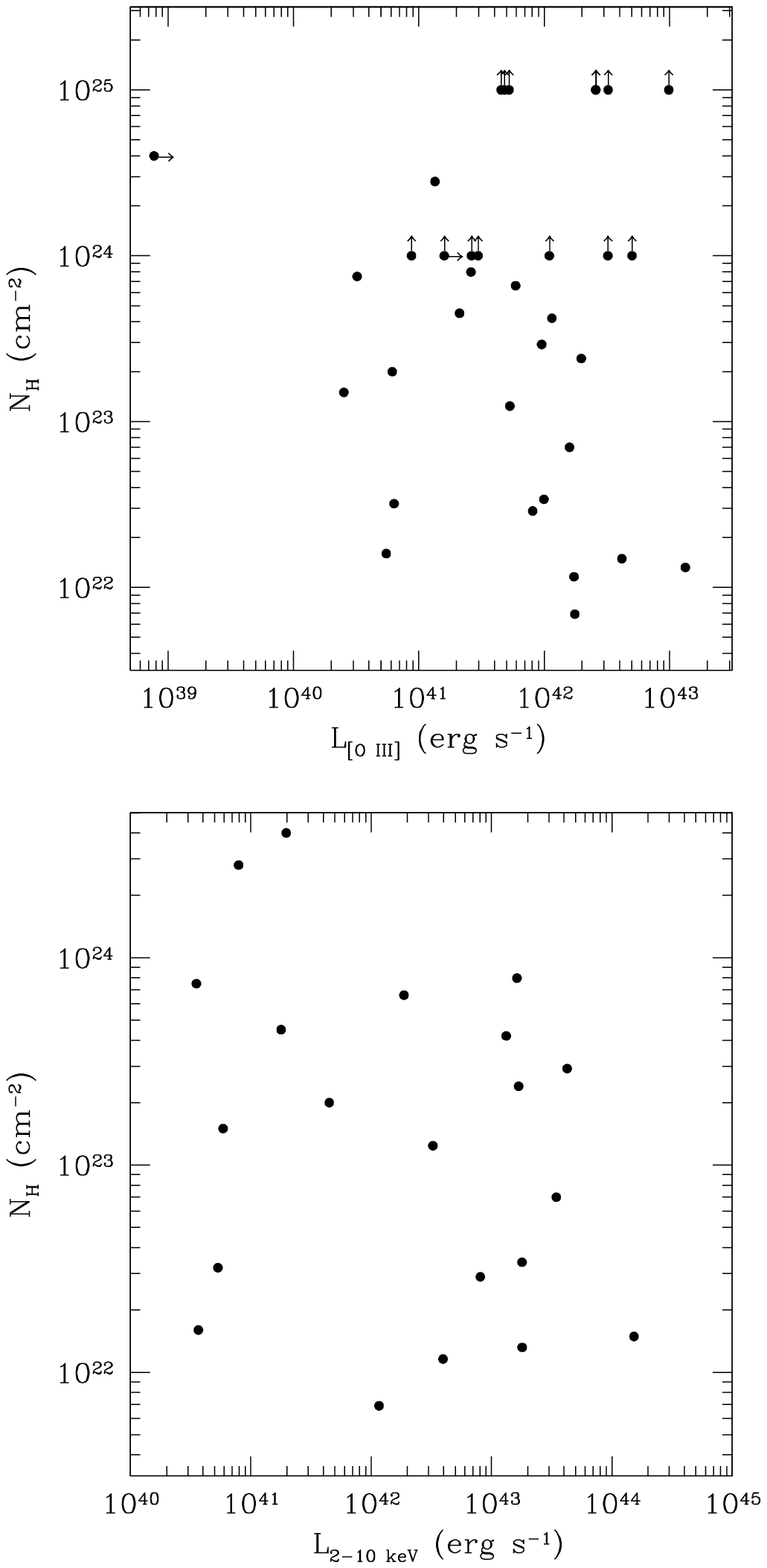]{The absorbing
N$_H$ as a function of the (extinction corrected) [OIII]
luminosity (upper panel) and of the absorption corrected 2--10 keV luminosity
(lower panel) for the subsample of 
objects with F$_{[OIII]} > 40\times 10^{-14}~ \rm cm^{-2}erg~s^{-1}$.
There is no evidence of a correlation between the
absorbing column density and the intrinsic luminosity of the AGN.
In the lower panel we only plot objects whose X-ray spectra show the direct
transmitted component, because it is not possible to estimate the 
intrinsic X-ray luminosity from the reflected component alone.}

\figcaption[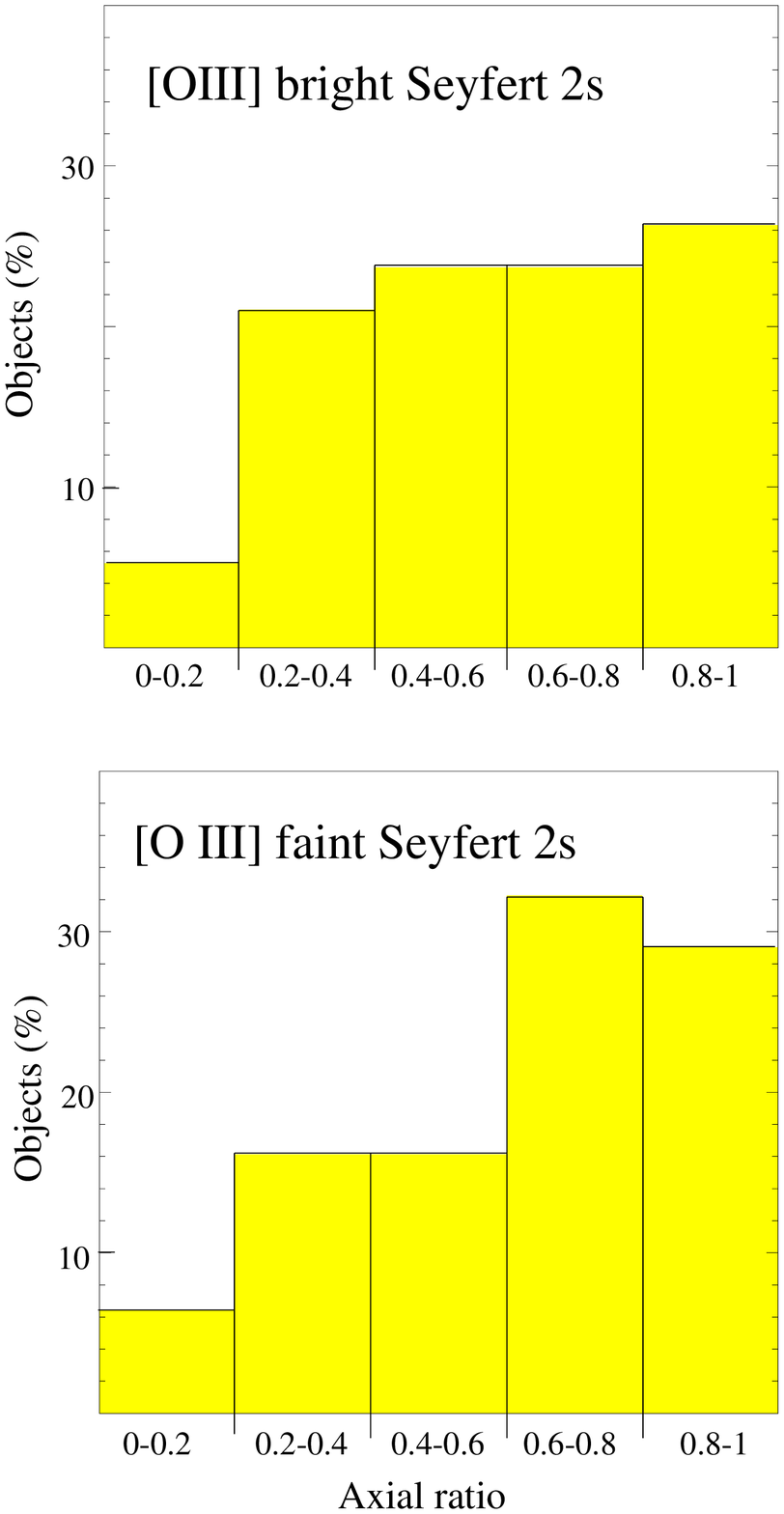]{The axial ratio distribution for the sources
with F$_{[OIII]} > 40$ erg cm$^{-2}$ s$^{-1}$ (upper histogram) and for
the sources with F$_{[OIII]} \le 40$ erg cm$^{-2}$ s$^{-1}$ (lower
histogram). We have not included elliptical and S0 galaxies. The fainter
subsample is clearly biased in favor of face--on systems.}

\figcaption[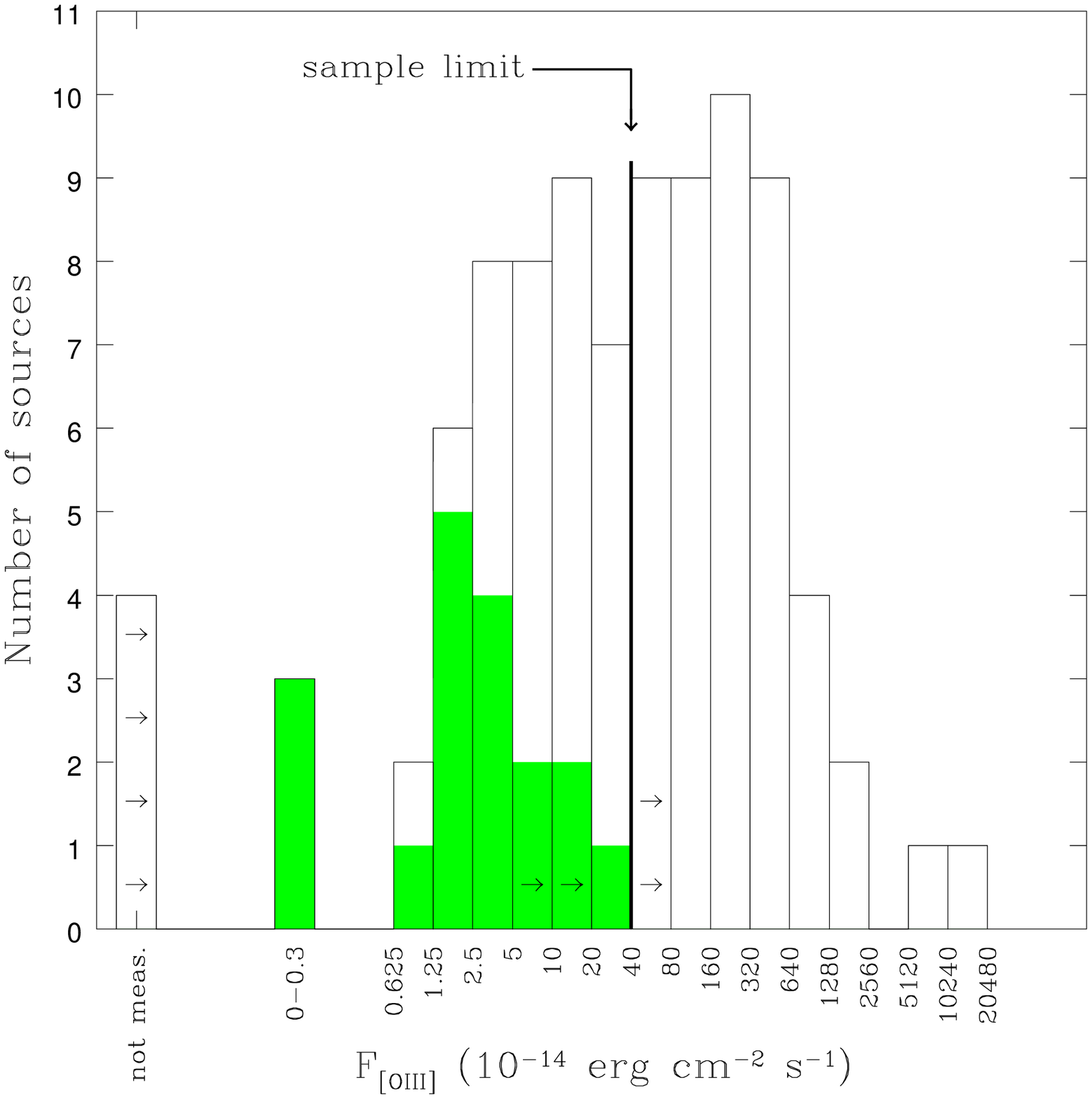]{The [OIII] flux distribution of the total sample
(Tab. 1 plus Tab. 2). The
shaded bars refer to the additional sources from Ho et al. (1997)}

\figcaption[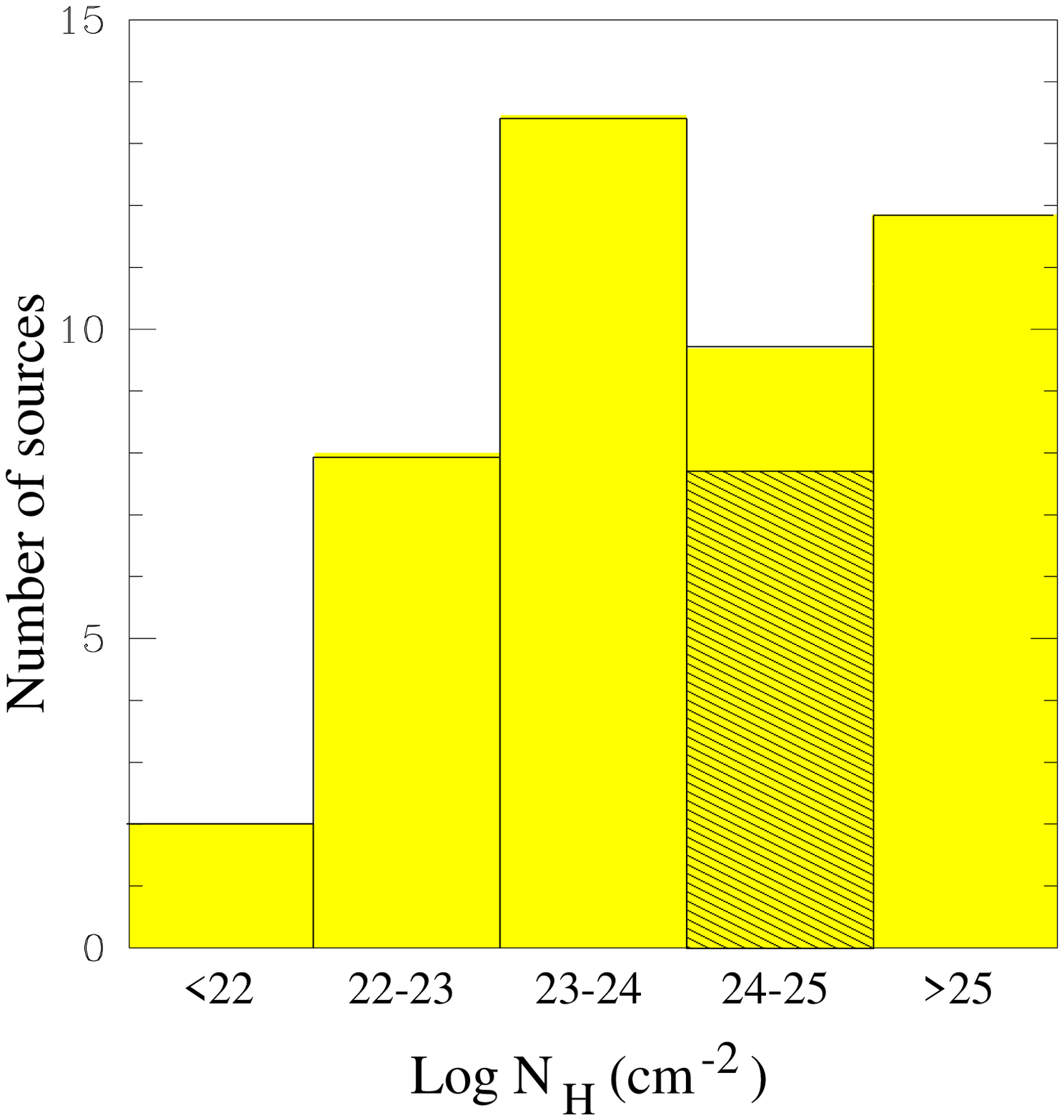]{The column density distribution of our final
subsample of 45
Seyfert 2 galaxies. The shaded part in the fourth bar indicates the
sources for which only a lower limit of 10$^{24}$ cm$^{-2}$ is available
at present.}

\figcaption[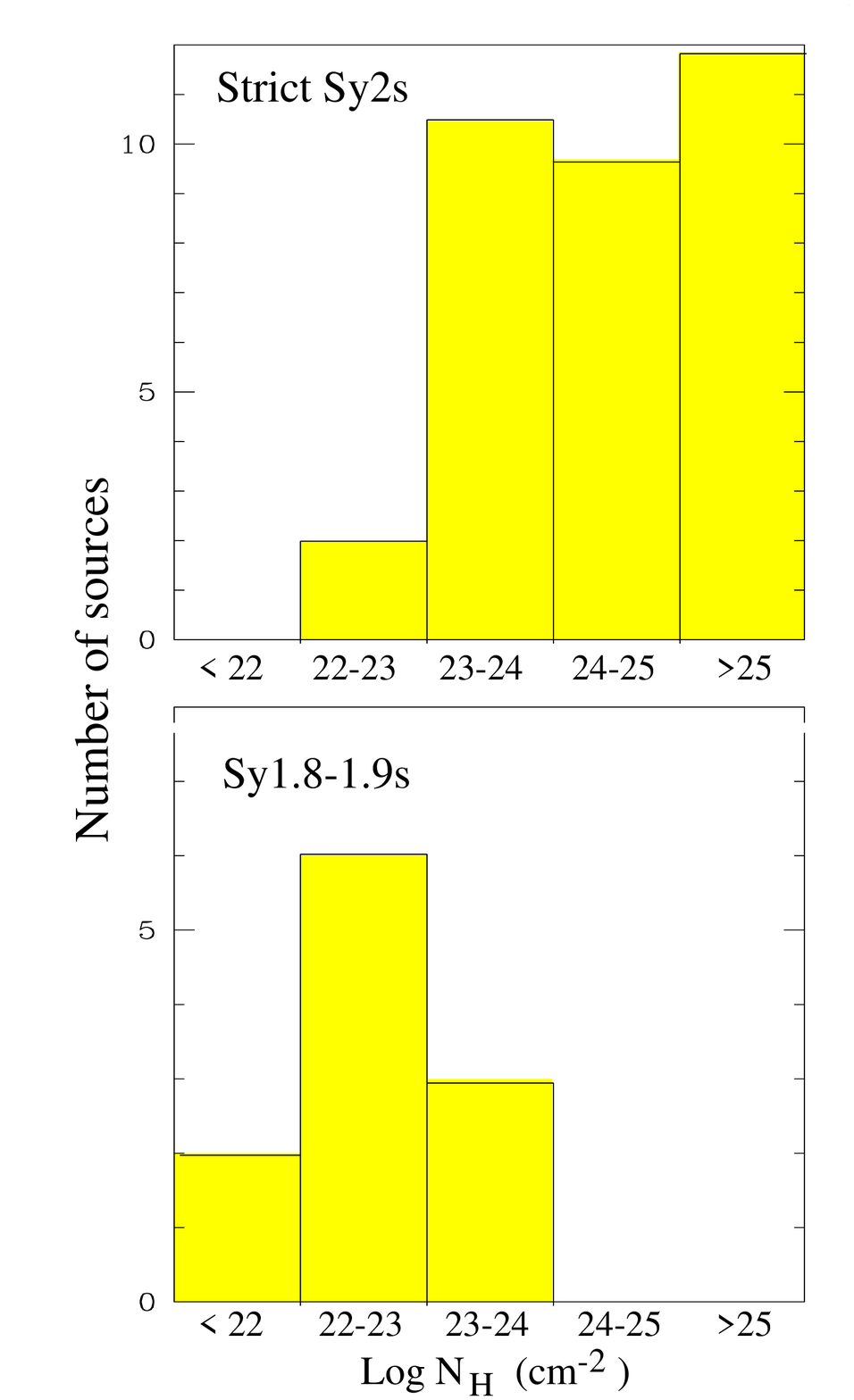]{The separate contribution to the N$_H$ distribution
from ``strict'' type 2 Seyferts (upper panel) and from intermediate
type 1.8--1.9 Seyferts (lower panel).}

\begin{figure}
\figurenum{1}
\epsscale{0.7}
\plotone{f1.eps}
\caption{}\label{f1}
\end{figure}

\begin{figure}
\figurenum{2}
\epsscale{0.7}
\plotone{f2.eps}
\caption{}\label{f2}
\end{figure}

\begin{figure}
\figurenum{3}
\epsscale{0.7}
\plotone{f3.eps}
\caption{}\label{f3}
\end{figure}

\begin{figure}
\figurenum{4}
\epsscale{1}
\plotone{f4.eps}
\caption{}\label{f4}
\end{figure}

\begin{figure}
\figurenum{5}
\epsscale{0.7}
\plotone{f5.eps}
\caption{}\label{f5}
\end{figure}

\end{document}